\magnification=\magstep1
\tolerance=500
\bigskip
\rightline{30 January, 2013}
\bigskip
\centerline{\bf Neutrinos and $v<c$}
\bigskip
\centerline{L.P. Horwitz$^{a,b,c}$ and I. Aharonovich$^b$}
\bigskip
\leftline{\it  ${}^a$ School of Physics, Tel Aviv University, Ramat Aviv
69978, Israel}
\smallskip
\leftline{\it ${}^b$ Department of Physics, Bar Ilan University, Ramat Gan
52900, Israel}
\smallskip
\leftline{\it ${}^c$ Department of Physics, Ariel University Center of
Samaria 40700, Israel}
\bigskip
\noindent{\it Abstract}
\smallskip
 
\input miniltx

\def\Gin@driver{dvips.def}
\input graphicx.sty

\par The Stueckelberg formulation of a manifestly covariant
relativistic classical and quantum mechanics is briefly reviewed and
it is shown that in this framework a simple (semiclassical)  model
exists for the description of neutrino oscillations. The model is shown
to be consistent with the field equations and the Lorentz force (developed here
without and with spin by canonical methods) for Glashow-Salam-Weinberg
 type non-Abelian fields interacting with the leptons. We discuss a
 possible fundamental mechanism, in the context of a relativistic
 theory of spin for (first quantized) quantum mechanical systems, for
 CP violation. The model also 
 predicts a possibly small ``pull back,'', {\it i.e.}, early arrival of a
 neutrino beam, for which the neutrino motion is almost everywhere
 within the light cone, a result which may emerge from future long
 baseline experiments
designed to investigate neutrino transit times with significantly
higher accuracy than presently available.
\bigskip
\noindent{\bf 1. Introduction}
\smallskip

\par In 1941 Stueckelberg [1] considered the possibility that in
classical mechanics, a timelike particle worldline can, due to
interaction, curve sufficiently in spacetime to evolve continuously
into a particle moving backward in time (see fig. 1). This
configuration was interpreted as representing particle-antiparticle
annihilation in classical mechanics. He parametrized this curve with
an invariant parameter $\tau$, which may be thought of as a world 
time\footnote{$^1$}{In 1973, Piron and
 Horwitz [2] generalized Stueckelberg's theory to be applicable to many
 body systems by assuming $\tau$ to be a universal world time
 essentially identical to
the time postulated by Newton. The classical Kepler problem was
solved in [2] by replacing the two body nonrelativistic potential
$V(r)$ by $V(\rho)$, where
$\rho$ is the invariant spacelike separation between the two
particles, and Arshansky and Horwitz [3] solved the quantum bound
state and scattering problems in this context; their method provides a
solution, in principle, for any two body problem described 
nonrelativistically by a
symmetric potential function $V(r)$.}
associated with the evolution. The time $t$ of the {\it event}
moving on a worldline of the type of fig. 1, corresponding to its
identification by Einstein as a coordinate of the Minkowski spacetime
manifold, is understood to be an observable in the same sense as the
space coordinates ${\bf x}$,{\it i.e.}, as the outcome of a {\it measurement},
 and the motion of an event in the eight
dimensional phase space $\{x^\mu,p^\mu\}$ is assumed to be governed
by a Hamiltonian dynamics. The Hamiltonian for a freely moving event
was taken by Stueckelberg to be
$$ K= {p^\mu p_\mu \over 2M}, \eqno(1)$$
where $M$ is an intrinsic property of the event with dimension of
mass; it may be thought of as the Galilean limiting mass of the
particle [4]. 
The quantity  $m^2 = -p^\mu p_\mu = E^2 -{\bf p}^2$ corresponds to
the actual measured mass of the associated particle. Since $t, {\bf x}$ are
dynamical variables, $E, {\bf p}$ are also, and the theory is
therefore intrinsically ``off-shell''. 
\par The Hamilton equations are,
quite generally,
$$ {\dot x}^\mu = {\partial K \over \partial p_\mu} \eqno(2)$$
and 
$$ {\dot p}_\mu= - {\partial K \over \partial x^\mu },\eqno(3)$$
where the dot corresponds to differentiation with respect to
$\tau$ (thus, for the free particle, 
 $d{\bf x} /dt = {\bf p} /E$, the standard relativistic 
definition of velocity). 
\par Stueckelberg [1] wrote a Schr\"odinger type equation to describe the
 evolution of the corresponding quantum state for a general
 Hamiltonian function of the quantum variables $x^\mu, p^\mu$ (which
 satisfy the commutation relations $[x^\mu,p^\nu]= i g^{\mu\nu}$, with
 metric signature $(-,+,+,+)$) 
$$ i {\partial \psi_\tau (x) \over \partial \tau} = K \psi_\tau (x),
\eqno(4)$$
where we denote $x\equiv (t,{\bf x})$. Since the wavefunction is
coherent in both space and time, the theory predicts interference in
time [5] in the same way that the nonrelativistic quantum
theory predicts interference in space.  An experiment of this type has
been done by Lindner {\it et al}[6] (this experiment was discussed
in the context of this theory by Horwitz[7]). The extension of the
theory  to describe particles with spin [8,9] introduces a foliation
of the quantum mechanical Hilbert
space as well as the corresponding quantum field theory as a
consequence of the spin-statistics relation[10] (Palacios
{\it et al}[11] have proposed an experiment which could be able to detect the
persistence of the entanglement of two particles with spin at different
times, as is consistent with this theory).
\par  This foliation,
which can be put into correspondence with the orientation of the
spacelike surfaces[10] forming the support of the complete set of local
observables providing the representations of quantum
fields[12], is involved in a fundamental way in $CP$
(or $T$) conjugation,as we shall discuss below, and therefore
intrinsically associated with the phenomenon of $CP$ violation 
(see, for example, refs. [13] for a discussion of neutrino flavor oscillations
and a comprehensive treatment of the $CP$ violation problem).  Similar 
considerations can be applied to
$CP$ and $T$ violations in the $K$, $B$ and $D$ systems [14]. There
have been many discussions in the framework of quantum field
theory[15], one of the earliest that of Weinberg[16], providing a
mechanism through coupling with a scalar field. It will be of
interest to investigate the relation between our 
observations in the framework of the first quantized theory with these studies.
\par Under a gauge transformation,
for which $\psi$ is replaced by $e^{i\Lambda}\psi$ (see,{\it e.g.},  Yang
and Mills [17]), this equation remains of the same form if one
replaces  $p^\mu$ by $p^\mu - e a^\mu$, where $e$ is here the coupling
to the 5D fields, and 
$ i {\partial \over \partial \tau}$ by 
$ i {\partial \over \partial \tau} + e a^5$, with $a^5$ a new scalar
field, and, under gauge transformation, the gauge compensation field
$a^\alpha$ is replaced by
$a^\alpha -\partial^\alpha \Lambda$, for $\alpha = 0,1,2,3,5$[18].  In
the following, we describe a
 model non-Abelian gauge theory, in which the fifth gauge field plays a
 fundamental role, for neutrino oscillations. We study the structure
 of a Lorentz invariant and gauge covariant Hamiltonian and Lagrangian
 for a particle in interaction with a non-Abelian gauge field, and the 
resulting field equations and Lorentz force, both without and with
spin. We point out mechanisms in this structure that can lead to $CP$
symmetry violation on a fundamental level.
\bigskip
 \noindent{\bf 2. The Model}
\par  In the flavor oscillations of the neutrino
system, interactions with the vector bosons of the Glashow-Salam-Weinberg (GSW)
theory[19] which induce the transition can produce pair
annihilation-creation 
events. In the framework of Stueckelberg theory, pair
annihilation and creation events can be correlated, as shown in
fig. 2, by following the world line.\footnote{$^2$}{ The
curve shown in fig. 2 should be thought of as corresponding to the 
expectation values computed with the local density matrix associated
with the gauge structured wave function of the neutrino beam.}
 The methods of Feynman's original paper, based on a spacetime
 picture[20], 
closely related to Stueckelberg's earlier formulation, would admit such a
construction as well. An ``on-shell'' version of our fig. 2 appears,
with sharp vertices, in the first of ref.[20].
\par It may be noted from this figure that there is a net decrease in the time
interval, possibly very small,  observed for the
particle to travel a certain distance. This mechanism is quite
different from 
that discussed by Floyd[21], Matone[22], and Faraggi[23], discussing the
phenomenon of early arrival as a purely quantum mechanical effect. The
most recent experiments[24] have shown that the arrival
times are consistent with light speed,{\it e.g.}, in the OPERA
experient [24], over the 732 km distance form CERN to Gran Sasso, an
arrival time of $6.5\pm 7.4 \big\{^{+8.3}_{-8.0}\big\} ns$  less than
 light speed arrival is reported, certainly consistent with light
speed.\footnote{$^3$}{It has been observed in the Supernova 1987a that
the neutrinos arrive about 3 hours before the light signal[25]. It has
been argued that the light is delayed, for example, by gravitational
effects on the virtual electron-positron pairs[26], but arguments of
Lorentz invariance[27] and the apparently universal applicability of the
geodesic behavior of light provide some difficulties for this view. On
the other hand, an advanced arrival of the order of 6.5 ns in each 730
km ( consistent with this data) would result in
approximately $3 \times 10^3$ hours early arrival.  However, as we
shall see below,
the mechanism for the oscillations associated with  such a ``pull-back''
involves the participation of the fifth field in an essential way, expected to
fall off far from sources. One may estimate on the basis of a 3 hour
early arrival the range of effectiveness of the fifth field, assuming
an advance of 6.5 ns in each 730 km where effective. A simple estimate
yields about 30 parsec(pc), as an effective size of the supernova. 
The Sun is about
$10^4$ pc from the center of the galaxy, so an effective range of
about 30 pc is not unreasonable. This argument is certainly not a proof
of a ``pull back'' ; it is meant to  show that a small effect of this
type could be consistent with the supernova 1987a data (see, moreover,
further discussion in [25]).}

\par Suppose, for example, that such oscillations can
occur twice during this transit [28] as in fig. 3. The particles (and
antiparticles) have  almost everywhere propagation
 speed less than light velocity (except for the vertices, which we
 estimate, based on the $Z,W$ lifetimes, to occur in about 
$10^{-22}$ seconds); it is clear from fig. 3 that an early arrival
would {\it not} imply, in this model, 
 that the neutrinos travel faster than light speed.  The effect noted
 by Glashow and Cohen [29],
indicating that \^Cerenkov radiation would be seen from faster than light 
neutrinos, would likely not be observed from the very short lived
vertices, involving interaction with the $W$ and $Z$ fields,
without sensitive detectors placed appropriately on the track
\footnote{$^4$}{ The neutrino arrivals detected at Gran Sasso appear to
be almost certainly normal particles. The ICARUS detector (Antonello
{\it et al} [30]) records no $\gamma$'s or $e^+ e^-$ pairs which would
be expected from \^Cerenkov radiation from
faster than light speed neutrinos. Our  model is consistent with the
presence of neutrinos for which the total travel time is closely
bounded by light velocity, since the ```pull-back'' effect can be very
small, limited (as in fig. 3) by the interval involved in the oscillation 
transitions.}
\par A quantum mechanical counterpart of this model, in terms of
 Ehrenfest wave packets, is consistent with this construction. The
 derivation of the 
 Landau-Peierls relation [31] $\Delta p \Delta t \geq {\hbar \over 2c}$
 in the framework of the Stueckelberg 
 theory[32] involves the assumption that the energy-momentum content of
 the propagating wave function contains predominantly components for which
 ${p\over E} <1$.  Interactions, {\it e.g.}, at the vertices of the
 curve in fig. 3,  can affect this distribution in such a
 way that, for some (small) interval of evolution, the wave packet can contain
 significant contributions to the expectation value of $p/E$ much
 larger than unity, and thus the dispersion $\Delta t$ in the
 Landau-Peierls relation can become very small without violating the
 uncertainty bound established by $<E/p>$. The interaction vertex may then
 be very sharp in $t$, admitting a precise manifestation of the
 deficit time intervals.
 \par The upper part of fig. 3 shows schematically the orbit of a
 neutrino in spacetime during its transit, according to this theory,
 in which the first( annihilation) event
 results in the transition from a $\nu_\mu$ to either a $\nu_\mu$ or
 $\nu_e$ through interaction with a GSW boson (for this simple
 illustration
 we consider only the $\mu$ and $e$ neutrinos, although there is no
 reason to exclude the $\tau$ neutrino) and the second
 (creation) event involves a transition from either of these
 states back to a $\nu_\mu,\nu_e$ state.

\bigskip
\noindent{\bf 3. Formulation of the  Non-Abelian Model}
\bigskip

\par The gauge covariant form of the Stueckelberg Hamiltonian, valid
 for the non-Abelian case as well as for the Abelian, with
 coupling $g$ to the 5D fields, is
$$K = {(p^\mu -gz^\mu)  (p_\mu - gz_\mu)\over 2M} - g z^5(x), \eqno(5)$$
where the $z^\mu$ fields are non-Abelian in the $SU(2)$ sector of the
electroweak theory.  Since, as we shall below, ${\dot x}^\mu$ is
proportional to $p^\mu -gz^\mu$, the local expectation of the square
of the ``proper time'' is proportional to that of the first term in the
Hamiltonian.   Therefore, see  we see that the local expectation of
$z^5$ must pass through that of the conserved value of $-K/g$ to admit
passage of the orbit through the light cone. In the
lower part of fig. 3, we have sketched a form for a smooth $z^5$ wave
(in expectation value) that
would satisfy this condition. Such a wave can be easily constructed as
the superposition of a few harmonic waves with different wavelengths
(originating in the spectral density of the neutrino wave functions
[see Eq. $(17)$]).
\par The occurrence of such a superposition can be understood from the
 point of view of the structure of the $5D$ GSW fields.  Working in
 the context of the first quantized theory, where the functions $\psi$
 belong to a Hilbert space $L^2(x,d^4x) \otimes d$, with $d$ the
 dimensionality of the gauge fields ($d=2$ corresponds to the
 Yang-Mills case [17] and the $SU(2)$ sector of the electroweak
 theory which we shall deal with here; our procedure for extracting
 the field equations and Lorentz force applies for any $d$), the
 field equations can be derived  
from the Lagrangian density (we consider the case of particles with spin
 in Sec. 5)
 $$ \eqalign{{\cal L} &= {1 \over 2}{\rm Tr}\bigl( i {\partial
 \psi\over \partial
 \tau}\psi^\dagger - i\psi{\partial \psi^\dagger\over \partial
 \tau}\bigr)\cr
 &-{1 \over 2M} {\rm Tr}\bigl[(p^\mu - gz^\mu)\psi((p_\mu -
 gz_\mu)\psi)^\dagger\bigr] \cr &+ g{\rm Tr}(z^5 \psi \psi^\dagger)
  - {\lambda \over 4} {\rm Tr} f^{\alpha \beta} f_{\alpha \beta}, \cr}
  \eqno(6)$$
where $\psi$ is a vector valued function representing
the algebraic action of the gauge field, and $\psi^\dagger$ is a
$2$-component (row) conjugate vector valued function; ${\cal L}$ is a
local scalar function. The operation Tr corresponds to a trace over
 the algebraic indices of the fields (the dimensional parameter 
$\lambda$ arises from
 the relation of these fields to the zero mode fields of the usual
 $4D$ theory [18]). For the variation of the field strengths we take
$\delta z^\alpha$ to be general infinitesimal Hermitian algebra-valued
 functions. Extracting the coefficients of these variations, with the 
definition of the non-Abelian gauge invariant field strength tensor[17] 
$$ f^{\alpha \beta} = \partial^\alpha z^\beta -\partial^\beta z^\alpha
-ig[z^\alpha, z^\beta], \eqno(7)$$
one obtains the field equations
$$\lambda \bigl[ \partial^\alpha f_{\beta\alpha} - ig[z^\alpha,
f_{\beta\alpha}] \bigr] = j_\beta  \eqno(8)$$
where
$$ j_\mu = {ig \over 2M}\{(\partial_\mu -igz_\mu) \psi \psi^\dagger
-\psi((\partial_\mu -igz_\mu)\psi)^\dagger\} ,\eqno(9)$$
and
$$ j_5 = g \psi \psi^\dagger\equiv \rho_5.\eqno(10)$$   
 
\par Let us now impose, as done by Yang and Mills
[17], the subsidiary condition
$$ \partial^\alpha z_\alpha = 0. \eqno(11)$$
We then obtain from $(8)$
$$ (-\partial_\tau^2 + \partial_t^2 - \nabla^2) z_\beta = j_\beta/\lambda
+ ig [z^\alpha, f_{\beta\alpha}],
\eqno(12)$$
where we have taken  the $O(4,1)$ signature for the fifth variable $\tau$.
Representing $z_\beta(x,\tau)$ in terms of its Fourier transfrom
$z_\beta(x,s)$, with
 $$ z_\beta(x,\tau) = \int ds e^{-is\tau} z_\beta(x,s), \eqno(13)$$
one obtains
$$ (s^2 + \partial_t^2 - \nabla^2) z_\beta(x,s)
 = j_\beta(x,s) / \lambda + ig \int d\tau e^{is\tau}
 [z^\alpha(x,\tau) , f_{\beta\alpha}(x,\tau)], \eqno(14)$$
providing a relation between the off-shell mass spectrum of the
$z_\beta$ field and the sources including the quantum mechanical
current as well as  the non-linear self-coupling of the fields. 
  \par Since the behavior of the $z_5$ field plays an essential role
  in the immediately applicable predictions of our model, consider the
  equation $(14)$ for $\beta=5$,
$$ (s^2 + \partial_t^2 - \nabla^2) z_5(x,s)
= j_5(x,s) / \lambda + ig \int d\tau e^{is\tau}
 [z^\nu(x,\tau) , f_{5\nu}(x,\tau)]. \eqno(15)$$
\par  In a zeroth approximation,
neglecting the nonlinear coupling term, we can study the equation
$$(s^2 + {\partial_t}^2 - \nabla^2) z_5(x,s) \cong
j_5 (x,s)/ \lambda. \eqno(16)$$
The source term is a convolution of the lepton wave functions in the
Fourier space, so that
$$(s^2 + {\partial_t}^2 - \nabla^2) z_5(x,s) \cong  {g \over
2\pi\lambda} \int ds'\psi(x,s') \psi^\dagger(x, s'-s). \eqno(17)$$
The Fourier representation over $s$ of the wave function corresponds to the
set of probability amplitudes for finding the particle in
the corresponding mass states; we expect these functions to peak in
absolute value, in free motion,  at the measured neutrino masses.
There is therefore the possibility of several mass values contributing
to the frequency of the spectrum of the $z_5$ field (the diagonal 
contributions contribute only to
its zero mode, a massless radiative field of essentially zero
measure). It is of interest to note that in order for the sources to
give rise to a form for the $z_5$ field of the type illustrated in
fig. 3, there must be at least three peaks in the mass distribution of
the wave functions, corresponding to three families of neutrinos.
This condition has been noted in a somewhat different context in the
last of [13] (p. 37)
and in other studies (for example refs.[33], discussing the three
family structure).
\vfill
\break

\bigskip
\noindent{\bf 4. The Non-Abelian Lorentz Force} 
\bigskip
\par We now turn to study the quantum trajectories of the particles
with non-Abelian gauge interactions to further check the consistency
of our model.  The Heisenberg equations of
motion are associated with expectation values for which the
classical motion is a good approximation if the wave packets are
fairly well localized. 
\par From the Hamiltonian $(5)$ one obtains
$$\eqalign{ {\dot x}^\lambda &= i [K,x^\lambda] \cr
&= {1 \over M} (p^\lambda - g z^\lambda), \cr} \eqno(18)$$
of the same form as the classical result.
\par  The second derivative is defined by
$$ {\ddot x}^\lambda = i[K, {\dot x}^\lambda] + {\partial {\dot
x}^\lambda \over \partial \tau}, \eqno(19)$$
where the last term is necessary because ${\dot x}^\lambda$ contains,
according to $(18)$, an explicit $\tau$ dependence which occurs in the
fields $z^\lambda$. One then obtains (the Lorentz force for the
non-Abelian case was also obtained, using an algebraic approach, in [34])

$$ {\ddot x}^\lambda=  - {g\over 2M} \{ {\dot x}^\mu, {f^\lambda}_\mu
\} - {g\over
M}f^{5\lambda} .   \eqno(20)$$
   \par Let us make here the crude approximation that was used in 
obtaining $(16)$,
{\it i.e.}, neglecting the nonlinear coupling to the spacetime components of
the field. Then, $(20)$ becomes, for the time component,
$$ {\ddot t} \cong -{g \over M} {\partial z^5 \over \partial
t}. \eqno(21)$$
\par The rising $z^5$ field (fig. 3), before the first passage through
the light cone,
would imply a negative curvature, as required. This consistency
  persists through the whole process. 
\par We further note that
$$ -{ds^2\over d\tau^2} = {2 \over M} (K + g z^5), \eqno(22)$$
so that
$$ { d\over d\tau} {ds^2\over d\tau^2} = -{2g \over M} {d z^5 \over
d\tau}, \eqno(23)$$
consistent as well with the form of fig. 3.
 \bigskip
\noindent{\bf 5.  The Hamiltonian for the Spin ${ 1 \over 2}$
Neutrinos}
\bigskip
\par The original construction of Wigner[35] for the description of
relativistic spin, for application to a relativistic quantum theory,
as explained in refs.[8] and [9], has a fundamental difficulty in that
the resulting  Wigner rotation [35] 
depends on momentum. The action of the resulting little group would depend on
this momentum vector and the expectation value of the
operator $x^\mu$, $<x^\mu>$ would not be covariant; $x^\mu$
acts as a derivative of the momentum and this would destroy the
unitarity of the little group action.  Inducing the representation on
a timelike four vector $n^\mu$ [8][9] preserves the covariance of $<x^\mu>$,
and also admits the possibility of linear superpositions over momenta
preserving the definition of the spin, for example, the construction of
wave packets in spacetime of definite spin. It also implies a
foliation of the Fock space for identical particles with strong
implications for many body systems as well as quantum field
theory[10].
\par  The wave functions, in
 coordinate or momentum space, now related in the usual way by Fourier
 transform, then transform , under a Lorentz transformation $\Lambda$, as
 $$ {\hat \psi}'_{\tau,\sigma}(x,n) = D(\Lambda,n)_{\sigma
 \sigma'}{\hat \psi}_{\tau, \sigma'}(\Lambda^{-1}x, \Lambda^{-1}n),
 \eqno(24)$$
 where $ D(\Lambda,n)$ is the associated Wigner rotation represented
 here in a fundamental representation of $SL(2,C)$. There are two inequivalent 
representations of $SL(2,C)$; since the operator
$\sigma^\mu p_\mu$ connects these two representations, and
such an operator would certainly occur in almost any dynamical theory, the
wave function must contain both fundamental representations. Within a
simple transformation (see {\it e.g.},[9],[10]), a vector combining
 the two forms of
 $L(n){\hat \psi}(x)$ (where the $SL(2,C)$ matrix $L(n)$ acts to bring 
$(-1,0,0,0)$ to $n^\mu$) of each type
 transforms like the Dirac wave function $\psi$, {\it i.e.},
$$ \psi'_\tau(x,n) = S(\Lambda) \psi_\tau(\Lambda^{-1}
x,\Lambda^{-1}n),  \eqno(25)$$
where $S(\Lambda)$ is generated in the usual way by 
$$ \Sigma^{\mu\nu} = {i \over 4}[\gamma^\mu, \gamma^\nu]. \eqno(26)$$
The norm, constructed of the sum of the norms of the two two-component
representations, is then given by\footnote{$^5$}{Note that in
 the Stueckelberg theory, the wave functions provide {\it local}
 probability amplitudes associated with particles, unlike the non-local
 properties pointed out by Newton and Wigner[36] of the Klein-Gordon or
 Dirac (on shell) functions[2].}

$${\Vert \psi\Vert_n}^2 = \int d^4 x {\bar \psi}_\tau(x,n) \gamma\cdot n
\psi_\tau(x,n) \eqno(27)$$
for each $n$ in the negative light cone. The complete positive
definite norm is given by the integral $\int {d^3{\bf n}\over n^0}$
over the full foliation (which is required to be convergent on the
full Hilbert space).
\par Following the method of ref.[9] for the non-Abelian case, we find
a Hamiltonian of the form
$$ K = {1 \over 2M} (p-gz)_\mu (p-gz)^\mu -{g \over 2M}f_{\mu\nu}
{\Sigma_n}^{\mu \nu} -gz^5, \eqno(28)$$
where 
 $${\Sigma_n}^{\mu\nu} = \Sigma^{\mu\nu} +K^\mu n^\nu - K^\nu n^\mu \equiv
{i \over 4} [\gamma_n^\mu, \gamma_n^\nu],
 \eqno(29)$$
with
$$ \gamma_n^\mu = \gamma_\lambda \pi^{\lambda \mu}_n; \eqno(30)$$
 the projection
$$ \pi^{\lambda \mu}_n = g^{\lambda\mu} + n^\lambda n^\mu \eqno(31)$$
 plays an important role in the description of the dynamics in the induced
representation. In $(28)$, the existence of projections on each index
in the spin coupling term implies that $f^{\mu\nu}$ can be replaced by
  ${f_n}^{\mu\nu}$ in this term, a tensor projected into the foliation 
subspace (see ref. [10] for discussion of the properties of this foliation).
\par  The $ {\Sigma_n}^{\mu\nu}$ generate a
Lorentz covariant form of the usual Pauli algebra (the compact $SU(2)$
part of the Lorentz algebra), and the $K^\mu$
generate the non-compact part of the Lorentz algebra [9] (since 
$n^\mu{\Sigma_n}^{\mu\nu}= K^\mu n_\mu = 0$, there are just three
independent $K^\mu$ and three ${\Sigma_n}^{\mu\nu}$). 
\par  The quantities $K^\mu$ and
$\Sigma_n^{\mu \nu}$ satisfy the commutation relations[9] 
 $$\eqalign{ [K^\mu,K^\nu] &= -i \Sigma_n^{\mu\nu}\cr
[\Sigma_n^{\mu\nu}, K^\lambda] &= -i[(g^{\nu\lambda} + n^\nu n^\lambda)
K^\mu - (g^{\mu\lambda} + n^\mu n^\lambda) K^\nu], \cr
[\Sigma_n^{\mu\nu}, \Sigma_n^{\lambda\sigma}] &= -i[(g^{\nu\lambda} +
n^\nu n^\lambda)\Sigma_n^{\mu\sigma} -(g^{\sigma\mu} + n^\sigma n^\mu)
\Sigma_n^{\lambda\nu} \cr
&-(g^{\mu\lambda} + n^\mu n^\lambda)\Sigma_n^{\nu\sigma} +
(g^{\sigma\nu} + n^\sigma n^\nu) \Sigma_n^{\lambda\nu}].\cr}
\eqno(32)$$
\par The last of
$(32)$ is the Lie algebra of $SU(2)$ in the spacelike surface
orthogonal to $n^\mu$. The three independent $K^\mu$ correspond to
the non-compact part of the algebra which, along with the
$\Sigma_n^{\mu\nu}$ provide a representation of the Lie algebra of the
full Lorentz group.  The covariance of this representation follows from
$$ S^{-1} (\Lambda) \Sigma_{\Lambda n}^{\mu\nu}S(\Lambda)
\Lambda_\mu^\lambda \Lambda_\nu^\sigma = \Sigma_n^{\lambda\sigma} . 
\eqno(33)$$
\par In the special frame for which  $n^\mu = (-1,0,0,0))$,
$\Sigma_n^{ij}$ become the Pauli matrices ${1\over 2} \sigma^k$
with $(i,j,k)$ cyclic, and $\Sigma_n^{0j} = 0$. In this frame there is
no direct electric type interaction with the spin in the minimal coupling
model $(28)$ (the theory admits a covariant form of electric coupling
of electric dipole type[9]; we will not consider this structure
here). We remark
that $\gamma^5$ commutes with this Hamiltonian, and therefore there is
a chiral decomposition (independently of the mass of the neutrinos)
that would admit the usual construction of the $SU(2)\times U(1)$
electroweak gauge theory.  The $SU(2)$ sector that we discuss below
would then apply to the left handed leptons.  The asymptotic (free)
solutions also admit a (foliated) helicity decomposition[9].
\par We record here the properties of the wave functions of a particle
with spin one half under the discrete symmetries $C,P$ and $T$,
obtained from the Stueckelberg-Schr\"odinger Eq. (4), with the
evolution generator $K$ given by $(5)$(computed for simplicity for
real Abelian gauge fields)[9]: 

$$\eqalign{ \psi_{\tau n}^C &= C \gamma^0 \psi_{-\tau n}^* (x)\cr
\psi_{\tau n}^P (x) &= \gamma^0 \psi_{\tau, -{\bf n}, n^0} (-{\bf x}, t),\cr
\psi_{\tau n}^T &= i \gamma^1\gamma^3 \psi_{-\tau,{\bf n}, -n^0}^* ({\bf
x}, -t), \cr
\psi_{\tau n}^{CP}(x) &= C \psi_{-\tau, -{\bf n}, n^0}(-{\bf x}, t)\cr
\psi_{\tau n}^{CPT}(x) &= i \gamma^5 \psi_{\tau, -n} (-{\bf x},-t),
 \cr}\eqno (34)$$
where $C= i\gamma^2 \gamma^0$.
  The $CPT$ conjugate wavefunction, according to its evolution in
  $\tau$, moves backwards in spacetime relative to the motion of
  $\psi_{\tau n}$. For a wave packet with  $E<0$ components, which
  moves backwards in $t$ as $\tau$ goes forward, it is the $CPT$
  conjugate wavefunction which moves forward with opposite charge, {\it i.e.},
  the observed antiparticle. We note that no Dirac sea is required for the
  consistency of the theory, since unbounded transitions to $E<0$ are
  prevented by conservation of $K$[9].
\par We shall discuss the possibilities of $CP$ violation provided by this
structure below.
\bigskip
\noindent{\bf 6. Lorentz force for spin $1/2$ particle with
non-Abelian gauge interactions} 
\bigskip
\par As in $(18)$, one obtains the particle velocity
$$ {\dot x}^\lambda= i [K, x^\lambda]= {1 \over M} (p^\lambda
-gz^\lambda). \eqno(35)$$
For the second derivative, from 
$(19)$ and $(32)$, we obtain
$$\eqalign{ {\ddot x}^\lambda &=-{g \over 2M} \{f^{\lambda\mu}, {\dot
x}_\mu\} - {g \over M} f^{5\lambda} \cr
&+ {g \over 2M^2} \partial^\lambda {f^n}_{\mu\nu} {\Sigma_n}^{\mu\nu} +
{ig^2 \over 2M^2} [{f^n}_{\mu\nu}, z^\lambda] {\Sigma_n}^{\mu\nu}.\cr}
\eqno(36)$$  
The third term of $(36)$ corresponds to a Stern-Gerlach type force.  Note
that we have included the subscript or superscript $n$ to the quantities
that are transverse in the foliation.
\par Under the assumption that the fields are not too rapidly varying,
and again neglecting coupling to the spacetime components of the field
$z^\alpha$, we see that 
the acceleration of the time variable along the orbit may again
be approximated by $(21)$. 

\bigskip
\noindent{\bf 7. Non-Abelian field equations with spin}
\bigskip
\par We are now in a position to write the Lagrangian for the full
theory with spin.
We take for the Lagrangian the form $(6)$ with an additional term for
the spin interaction and factors of $\gamma^0(\gamma\cdot n)$ to
assure covariance, yielding under variation of
$\psi^\dagger$ the
Stueckelberg equation for $\psi$ with Hamiltonian $(28)$:
$$\eqalign{ {\cal L}_n &= {1 \over 2}{\rm Tr}\bigl( i {\partial\psi
\over \partial
 \tau}{\bar \psi} - i\psi{\partial {\bar\psi}\over \partial
 \tau}\bigr)(\gamma\cdot n)\cr
 &-{1 \over 2M} {\rm Tr}\bigl[(p^\mu - gz^\mu)\psi\overline{(p_\mu -
 gz_\mu)\psi)}(\gamma \cdot n)\bigr] \cr
&+ g{\rm Tr}(z^5 \psi {\bar\psi}(\gamma \cdot n))
  - {\lambda \over 4} {\rm Tr} f^{\alpha \beta}
 f_{\alpha \beta}\cr
&+ {g \over 2M} {\rm Tr}(f_{\mu\nu}
 {\Sigma_n}^{\mu\nu} \psi {\bar \psi}(\gamma\cdot n)).\cr} \eqno(37)$$
 Defining $j_\alpha$ as in $(9),(10)$, but with the factor 
$\gamma^0\gamma\cdot n$,
required for covariance, {\i.e.}, 
$$ j_{n\mu} = {ig \over 2M}\{(\partial_\mu -igz_\mu) \psi{\bar \psi}
-\psi \overline{(\partial_\mu -igz_\mu) \psi)}\}(\gamma\cdot
n), \eqno(38)$$
and
$$ j_{n5} = g \psi {\bar \psi}(\gamma \cdot n)\equiv \rho_n,\eqno(39)$$
the variation of the Lagrangian 
with respect to the $z$-fields , where we have used the cyclic properties of
matrices under the trace, yields, setting the coefficients of
$\delta z^\nu,\delta z^5$ equal to zero, the field equations
$$ \lambda (\partial_\beta f^{5\beta} -ig[z_\beta, f^{5\beta}]) = \rho_n
\eqno(40)$$ 
and
$$ \eqalign{\lambda(\partial_\beta f^{\nu\beta} &-ig [z_\beta, f^{\nu
\beta}])\cr  &={j_n}^\nu + {g\over M} {\Sigma_n}^{\mu\nu} \bigl\{
\partial_\mu \rho_n -ig[z_\mu, \rho_n]\bigr\}. \cr} \eqno(41)$$
  Eq. $(41)$ corresponds to a
Gordon type decomposition of the current, here projected into the
foliation space (spacelike) orthogonal to $n^\mu$.  Note that the covariant
derivative of $\rho_n$ in the last term is also projected into the
foliation space.
\par With the subsidiary condition $\partial^\beta z_\beta =0$, as
before, we may write the field equations as
    $$ \lambda(-\partial^\beta \partial_\beta z^5 -ig[z_\mu, f^{5\mu}]) =
    \rho_n \eqno(42)$$ 
and
$$ \lambda(-\partial^\beta \partial_\beta  z^\nu  
-ig [z_\beta, f^{\nu\beta}])  
= {j_n}^\nu + {g\over M} {\Sigma_n}^{\mu\nu} \bigl\{
\partial_\mu \rho_n -ig[z_\mu, \rho_n]\bigr\}. \eqno(43)$$
Note that the spin coupling is not explicit in $(42)$. Neglecting, as before, 
coupling to the spacetime components,  
one reaches the same conclusions for the approximate behavior of the
$z^5$ field, {\it i.e.}, as determined by Eq.$(17)$ with $\psi^\dagger$
replaced by  ${\bar\psi}\gamma \cdot n$.  The latter reduces to the
 same expression for $n^\mu \rightarrow (-1,0,0,0)$.
.\bigskip
\noindent{\bf 8. $CP$ and $T$ Conjugation.}
\bigskip
\par The association of this timelike vector with the spacelike
surfaces used by Schwinger and Tomonaga[12] for the quantization of
field theories has been recently discussed[10]. These spacelike
surfaces form the
support of a complete set of commuting local observables on which the
Hilbert space of states in constructed. It follows from the above properties
of the wave functions for a particle with spin, that the $CPT$ conjugate
theory would be associated with the same spacelike surface,
corresponding to $\pm n^\mu$. However, the $CP$ conjugate, taking
${\bf n} \rightarrow -{\bf n}$ and $n^0 \rightarrow n^0$ refers to an
entirely different spacelike surface (the time reversed states, for
which ${\bf n} \rightarrow {\bf n}$ and $n^0 \rightarrow -n^0$ are
associated with this spacelike surface as well, with reflected unit timelike
vector). The equivalence of the physical processes described in these
two frameworks would depend on the existence of an isometry (including
both unitary and antiunitary transformations) changing the basis of
the space from the set of local
observables on the first spacelike surface to those defined on the
conjugated surface as well as the equivalence of the physics evolving
from it after the $CP$ (or $T$) conjugation. 
\par The spin coupling term in $(28)$
contains the possibility of $CP$ violation in generating a physics that
is inequivalent on the new spacelike surface. The nonrelativistic
quantum theory with Zeeman type ${\bf \sigma}\cdot {\bf H}$
coupling is, of course, not invariant under $T$ conjugation.
Precisely the same situation is true in the corresponding relativistic
equation $(28)$; as we have pointed out, in the special frame in
which $n^\mu = (-1,0,0,0)$, the matrices $\Sigma_n^{\mu\nu}$ reduce to
Pauli matrices. Under Lorentz transformation they still generate the
algebra of $SU(2)$ in a fundamental representation, and therefore still
contain the imaginary unit. Therefore, the physical evolution on the
$CP$ conjugate spacelike surface is not, in general, equivalent to the 
original evolution. For this phenomenon to occur, it is necessary that
there be present an $f_{\mu\nu}$ field.  In addition to
self-interaction effects, for which the intrinsic $CP$ violation can be
expected to cancel, the Stueckelberg oscillation diagram of fig.2
 suggests the existence of fields present in the equations of
motion of the second branch due to the proximity of the accelerated
motion in the first branch, thus providing a fundamental mechanism for
$CP$ violation.  A consequence of this structure is that the physics
in the corresponding $CP$ conjugated system of the quantum fields,
evolving from the $CP$ conjugate spacelike surface, could be
inequivalent. 
\bigskip
\noindent{\bf Conclusions}
\bigskip
\par We have argued that, according to the derivation of the Landau-Peierls
relation given in ref.[32], the vertices of the
neutrino-antineutrino transitions may be very sharp, and provide for a
rather precise ``pull back'' of the time interval.  Significantly
higher precision than available in the present experiments would be
necessary to see such an effect. 
\par We have worked out the equations describing the Lorentz forces
and the field equations of
the corresponding ($5D$) non-Abelian gauge theory, with the help of
Stueckelberg type Hamiltonians both for the spinless case
 and for the case of relativistic particles with spin in interaction
with such a nonabelian gauge field, and have shown that the conclusions reached
 are, in lowest approximation, consistent with our simple model. We
 emphasize that, in the framework of the Stueckelberg model, the
 dynamics of the fifth gauge field, modulated by the particle mass
 spectrum contained in the wave function (as in Eq.$ (17)$), plays an 
essential role for the oscillation process.
\par The presence of spin, described in the
relativistic framework of Wigner[35], as in refs. [9][10], introduces
a foliation in the
Hilbert space and in the structure of the fields, both classical and
quantum. Since, in
Tomonaga-Schwinger quantization of the fields, the spacelike surface
constructed to define a complete set of local observables is
characterized by being orthogonal to the timelike vector $n$ of the
foliation[10], the actions of the discrete $CP$ or $T$ transformations
change the basis for the construction of the Hilbert space to essentially
different spacelike surfaces. Along with the form of the spin
coupling term in $(28)$, this suggests a model for $CP$ or $T$
violation on the first quantized level.  Further
consequences of this foliation will be explored elsewhere.
\par We furthermore remark that our model would be applicable to the $K$,
$B$ and $D$ systems[14] as well, manifested by the quark gluon interactions in
their substructure. This possibility is under study.
\vfill
\break

\bigskip
\noindent{\it Acknowledgements}
\bigskip
\par We wish to thank 
Asher Yahalom, Jacob Levitan, Avi Gershon, Andrew Bennett and David
Owen for discussions.   We thank
E.R. Floyd for bringing his work to our attention, and Steve Weinberg
for a communication on the 1987a supernova.  We are also grateful 
to the Physics
Department at the Ben Gurion University of the Negev, and the Physics
Department at Bar Ilan University for opportunities
to give seminars on this work in its early stages. We would also like to
thank Itzhak Rabin for his encouragment and Faina 
Shikerman for designing the figures and for discussions. 

\bigskip
\noindent{\it References}
\frenchspacing
\bigskip
\item{[1]} Stueckelberg ECG, Helv. Phys. Acta {\bf
14}, 322-323,588-594 (1941); {\bf 15}, 23-27 (1942).
\item{2.} Horwitz LP and  Piron C, Helv. Phys. Acta {\bf 46,} 316-326 (1973).
\item{3.}  Arshansky RI and Horwitz LP Jour. Math. Phys. {\bf 30} 66,380
(1989).
\item{4.}  Horwitz LP. Schieve WC and  Piron C,  Ann. Phys.{\bf 137},
306-340 (1981). 
 \item{5.} Horwitz LP and Rabin Y, Lett.al Nuovo Cim {\bf 17} 501
 (1976).
\item{6.} Lindner F, {\it et al}  Phys. Rev. Lett.  {\bf 95}, 040401
(2005).
\item{7.} Horwitz L, Phy. Lett. A {\bf 355}, 1-6 (2006).
\item{8.} Horwitz L, Piron C and Reuse F,
 Helv. Phys. Acta  {\bf 48}, 546 (1975)
\item{9.}  Arshansky R.and Horwitz LP,
 J. Phys. A: Math. Gen. {\bf 15}, L659-662 (1982).
\item{10.} Horwitz LP, Jour. of Phys.A: Math. Theor. {\bf 46} 035305
(2013).
\item{11.} Palacios A, Rescigno TN, and McCurdy CW,
Phys. Rev. Lett. {\bf 103}, 253001-1-253001-4 (2009).
\item{12.} Tomonaga S, Prog. Theor. Phys.{\bf 1}, 27 (1946), Schwinger
J, Phys. Rev. {\bf 74} 1439 (1948).See also Weiss P,
Proc. Roy. Soc. {\bf A169},102 (1938), Jauch JM and Rohrlich F. {\it
The Theory of Photons and Electrons}, Springer-Verlag (1976).  
\item{13.}  Bilenky SM and
 Pontecorvo B, Phys. Rep. (Phys. Lett. C){\bf 41}, 225 (1978);
 Nunokawa H, Parke S and Valle WFJ, Prog. Part. Nucl. Phys. {\bf 60},
 338 (2008), arXiv:0710.0554 (2007).
\item{14.} Kayser B, {\it The frequency of neutral meson and neutrino
oscillation}, SLAC-PUB-7123 (2005);{\it Neutrino Physics} SLAC Summer
Institute on Particle Physics, Aug. 2-13 (2004); Aaltonen T {\it et
al}, Phys. Rev. D{\bf 85}, 012009 (2012).See also, Lees JP {\it et
al}, Phys. Rev. Lett. {\bf 109}, 211801 (2012), arXiv:1207.5832;
Bhattacharya B, Gronau M, and Rosner JL, Phys. Rev. D {\bf 85} 054014
(2012), arXiv:1207.0761.
\item{15.} For example, Sassaroli E,{\it Neutrino flavor mixing and
oscillation in Field theory}, arXiv:hep-ph/9805480 (1998); see also 
Fayyazuddin and Riazuddin, {\it  Modern
Introduction to Particle Physics}, World Scientific, Singapore (2012)
and Mohapatra RN, {\it Unification and Supersymmetry}, Springer, New
York (1986).
\item{16.} Weinberg S, Phys. Rev. Lett. {\bf 37} 657(1976); see also
Peccei RD and Quinn HR, Phys. Rev. D {\bf 16}, 1791 (1977).
\item{17.} Yang CN and Mills R, Phys. Rev. {\bf 96}, 191-195 (1954).  
\item{18.} Saad D, Horwitz LP and Arshansky RI, Found. Phys. {\bf
19}, 1125-1149 (1989). See also,  Aharonovich I and Horwitz LP,
Jour. Math. Phys. {\bf 47}, 122902-1-12290226 (2006); {\bf 51}, 
052903-1-052903-27 (2010); {\bf
52}, 082901-1-08290111 (2011); {\bf 53}, 032902-1-032902-29 (2012); 
Eur. Phys. Lett. {\bf 97}, 60001-p1-60001-p3 (2012).
\item{19.} Glashow S,  Iliopoulis J, and Maini L, Phys. Rev. {\bf
D2}, 1285-1292 (1970); Salam A,  {\it Elementary Particle Theory:
Relativistic Groups and Analyticity}, 8th Nobel Symposium, 367-377,
ed.  Svartholm N, Almquist and Wiksell, Stockholm (1968);
 Weinberg S, Phys. Rev. Lett.{\bf 19}, 1264-1266 (1967).
\item{20.} Feynman RP, Phys. Rev. {\bf 76}, 749-784, 769-789 (1949).
See also, Schwinger J, Phys. Rev. {\bf 82}, 664-679 (1951).
\item{21.} E.R. Floyd, {\it Opera superluminal neutrinos per quantum
trajectories} arXiv: 1112.4779 (2012). 
\item{22.} M. Matone, {\it Superluminality and a curious phenomenon
in the relativistic Hamilton-Jacobi equation} arXiv: 1109.6631 (2011).
\item{23.} A. Faraggi, {\it Opera data and the equivalence postulate
of quantum mechanics}, arXiv:1110.1857 (2011).
\item{24.}Adam T {\it et al},Jour. of High Energy Physics {\bf 10} 093
(2012), arXiv:1109.4897; Agafonova N Yu {\it et al},
      Phys. Rev. Lett. {\bf 109} 070801 (2012) arXiv:1208.1392.
\item{25.}Arnett WD, Bahcall JN, Kirschner RP, and Woosley SE,
Ann. Rev. Astron. Astrophys {\bf 27}, 629 (1989).
\item{26.} Franson JD, {\it Reduced photon velocities in the OPERA
neutrino experiment and Supernova 1978a}, arXiv:1111.6986 (2012).
\item{27.} Personal communication,  Weinberg S.
\item{28.} See, for example, Valle JWF,  Journal of Physics: 
Conference Series {\bf 56}, 473-505 (2006).
\item{29.} Cohen AG and Glashow SL,  arXiv:1109.6562 (2011).
\item{30.} Antonello M,  {\it et al},  arXiv: 1110.3763 (2011).
\item{31.} Landau L and Peierls R, Z. Phys. {\bf 69},
56-69 (1931).
\item{32.} Arshansky R and Horwitz LP, Found. of Phys. {\bf 15},
701-715 (1984).
\item{33.}For example, Fogli GL, Lisi E and Scioscia G, {\it
Accelerator reactor neutrinos oscillation experiments in a simple
three-generation framework}, IASSNS-AST 95/28, BARI-TH/219-95, arXiv
hep-ph/9506350 (1995); Bandyopadahyay A,Choubey S, Goswami S and
Kamales K, Phys.Rev. D{\bf 65} 073031 (2002) and arXiv:hep-ph/0110307 (2002). 
See also ref [14].
\item{34.} Land MC, Shnerb N and Horwitz LP, 
Jour. Math. Phys.{\bf 36}, 3263-3288 (1995).
\item{35.}Wigner E,  Ann. of Math. {\bf 40}, 149-204 (1939).
\item{36.} Newton TD and Wigner E, Rev. Mod. Phys. {\bf 21},
400-406(1949).

\vfill
\break

\includegraphics[width=12cm]{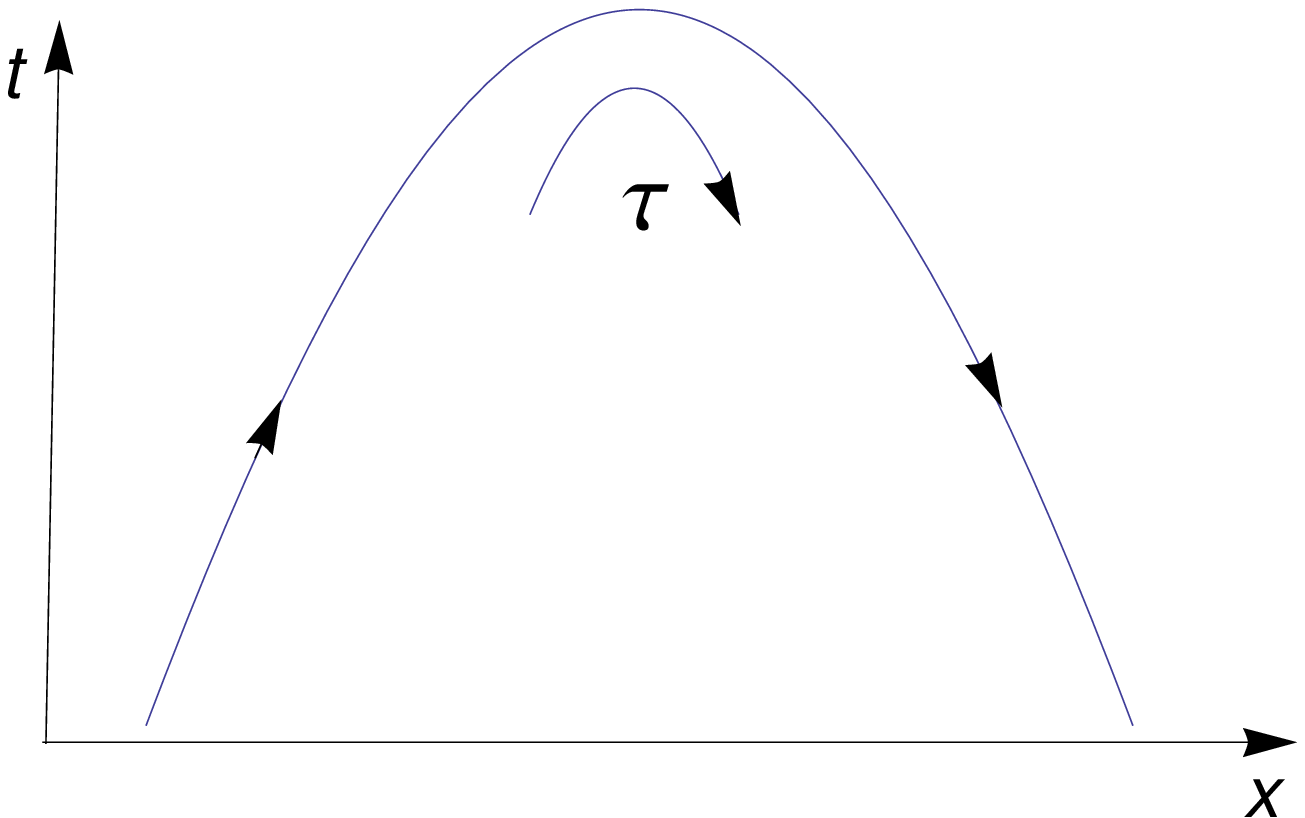}
 \smallskip
\centerline{Figure 1: A particle turning backward in $t$ appears as a particle anti-particle annihilation}
 \bigskip
\includegraphics[width=12cm]{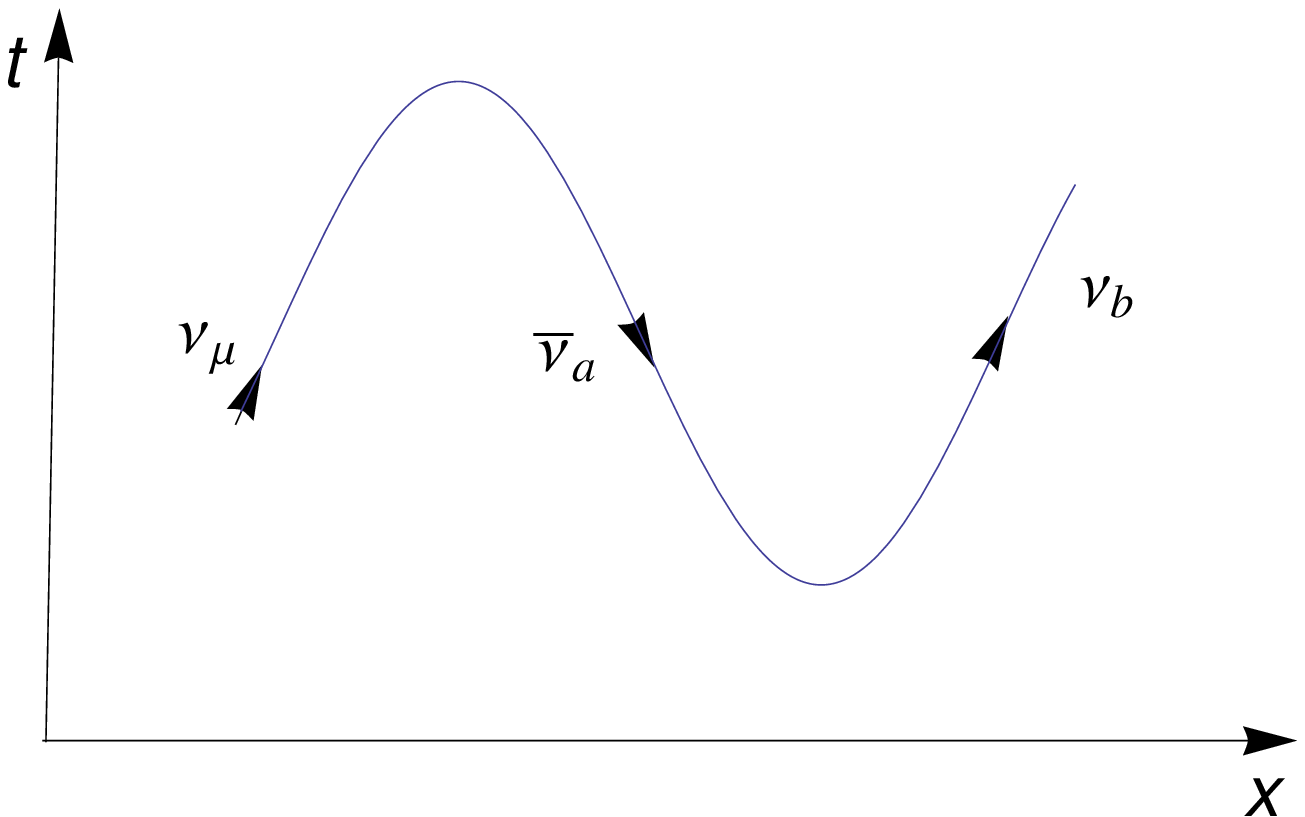}
 \smallskip
 \centerline{Figure 2: Pair annihilation seen as a shorter transit time.}
 
 \bigskip
\includegraphics[width=12cm]{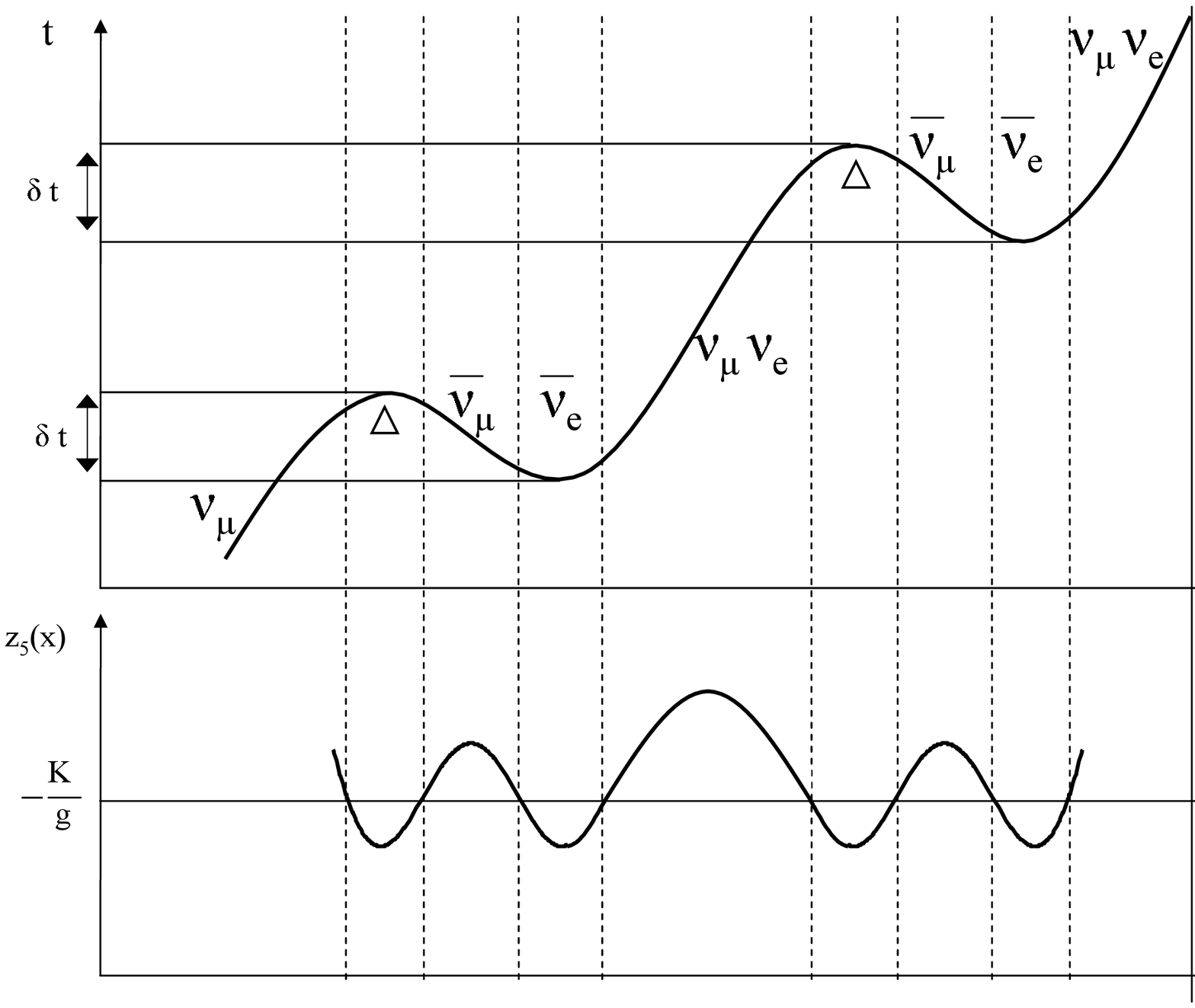}
 \smallskip
 \centerline{Figure 3: Neutrino travel model with shorter transit time.}

\end